\documentclass[aps,prl,twocolumn,groupedaddress,superscriptaddress,floatfix]{revtex4}

\usepackage[utf8]{inputenc}
\usepackage[T1]{fontenc}
\usepackage{slashed}
\usepackage{times}

\usepackage{amssymb,amsfonts,amsmath,amsthm}

\usepackage{slashed}
\usepackage{dcolumn}
\newcolumntype{l}[1]{D{.}{\cdot}{#1}} 

\usepackage{graphicx}
\usepackage[tight,hang,raggedright,normalsize]{subfigure}

\usepackage[
            pdftitle={Electroproduction of the N*(1535) resonance at large momentum transfer},
            pdfauthor={Vladimir Braun},
            pdfsubject={Electroproduction of N*(1535) resonance},
            pdfkeywords={moments, hadron, nucleon, baryon, quark, distribution,wave,function},
            pdfpagemode=UseOutlines,
            colorlinks=true,
            pdfproducer={University of Regensburg},
            ]{hyperref}

\newcommand{\up}{{\uparrow}}
\newcommand{\down}{{\downarrow}}

\begin{document}

\preprint@{
\hspace*{-1mm}DESY 09-025\\
\hspace*{2.3mm}Edinburgh 2009/02\\
\hspace*{2.3mm}LTH 822\\
}
\vspace{2mm}

\title{Electroproduction of the $N^*(1535)$ resonance at large momentum transfer}

\author{V.~M. \surname{Braun}}
    \affiliation{Institut f\"ur Theoretische Physik, Universit\"at Regensburg, 93040 Regensburg, Germany}
\author{M. \surname{G\"ockeler}}
    \affiliation{Institut f\"ur Theoretische Physik, Universit\"at Regensburg, 93040 Regensburg, Germany}
\author{R. \surname{Horsley}}
    \affiliation{School of Physics and Astronomy, University of Edinburgh, Edinburgh EH9~3JZ, UK}
\author{T. \surname{Kaltenbrunner}}
    \affiliation{Institut f\"ur Theoretische Physik, Universit\"at Regensburg, 93040 Regensburg, Germany}
\author{A. \surname{Lenz}}
    \affiliation{Institut f\"ur Theoretische Physik, Universit\"at Regensburg, 93040 Regensburg, Germany}
\author{Y. \surname{Nakamura}}
    \affiliation{Institut f\"ur Theoretische Physik, Universit\"at Regensburg, 93040 Regensburg, Germany}
    \affiliation{Deutsches Elektronen-Synchrotron DESY, 15738 Zeuthen, Germany}
\author{D. \surname{Pleiter}}
    \affiliation{Deutsches Elektronen-Synchrotron DESY, 15738 Zeuthen, Germany}
\author{P.~E.~L. \surname{Rakow}}
    \affiliation{Theoretical Physics Division, Department of Mathematical Sciences, University of Liverpool, Liverpool L69~3BX, UK}
\author{J. \surname{Rohrwild}}
    \affiliation{Institut f\"ur Theoretische Physik, Universit\"at Regensburg, 93040 Regensburg, Germany}
\author{A. \surname{Sch\"afer}}
    \affiliation{Institut f\"ur Theoretische Physik, Universit\"at Regensburg, 93040 Regensburg, Germany}
\author{G. \surname{Schierholz}}
    \affiliation{Institut f\"ur Theoretische Physik, Universit\"at Regensburg, 93040 Regensburg, Germany}
    \affiliation{Deutsches Elektronen-Synchrotron DESY, 22603 Hamburg, Germany}
\author{H. \surname{St\"uben}}
    \affiliation{Konrad-Zuse-Zentrum f\"ur Informationstechnik Berlin, 14195 Berlin, Germany}
\author{N. \surname{Warkentin}}
    \affiliation{Institut f\"ur Theoretische Physik, Universit\"at Regensburg, 93040 Regensburg, Germany}
    \email{nikolaus.warkentin@physik.uni-regensburg.de}
\author{J.~M. \surname{Zanotti}}
    \affiliation{School of Physics and Astronomy, University of Edinburgh, Edinburgh EH9~3JZ, UK}

%\date{}

\begin{abstract}
We report on the first lattice calculation of light-cone distribution
amplitudes of the $N^*(1535)$ resonance, which are used to
calculate the transition form factors at large momentum transfers
using light-cone sum rules. In the region  $Q^2 > 2$~GeV$^2$, where the
light-cone expansion is expected to converge, the results appear to
be in good agreement with the experimental data.
\end{abstract}

% 12.38.Gc Lattice QCD calculations
% 12.38.Lg Other nonperturbative calculations
% 13.40.Gp 	Electromagnetic form factors
% 14.20.Gk 	Baryon resonances with S=0



\maketitle

\phantomsection
\addcontentsline{toc}{subsubsection}{Introduction.}
\paragraph{Introduction. ---}
Electroproduction of nucleon resonances has long been recognized as 
an important tool in the exploration of the nucleon structure at different scales.
Transition form factors to nucleon excited states allow one to study
the relevant degrees of freedom, the wave function and the interaction 
between the constituents.
Quantum chromodynamics (QCD) predicts  
\cite{Chernyak:1977as,Chernyak:1977fk,Efremov:1978rn,Efremov:1979qk,Lepage:1979za,Lepage:1980fj}
that at large momentum transfer the form factors become increasingly dominated by the 
contribution of the valence Fock state with small transverse separation between the partons. 
There is a growing consensus that
perturbative QCD (pQCD) factorization based on hard gluon exchange
is not reached at present energies; however, the emergence of quarks and gluons 
as the adequate degrees of freedom is expected to happen earlier, 
at $Q^2\sim$ a few GeV$^2$. In this transition region the form factors  can be measured to 
high accuracy, see e.g. \cite{Burkert:2008mw}, and such measurements are in fact part of the 
experimental proposal for the 12 GeV upgrade at Jefferson Lab \cite{Nstar_workshop}.

Theoretical progress in the QCD description of transition form factors has been slow. 
The major problem is that any attempt at a quantitative description of form factors 
in the transition region must include soft nonperturbative contributions which 
correspond to the overlap integrals of the soft wave functions, 
see, e.g., \cite{Kroll:1995pv,Bolz:1996sw}.
In particular, models of generalized parton distributions (GPDs)
usually are chosen such that the experimental data on form factors 
are described by the soft contributions alone, 
cf.~\cite{Belitsky:2003nz,Diehl:2004cx,Guidal:2004nd}. 
A subtle  point for these semi-phenomenological 
approaches is to avoid double counting of hard rescattering 
contributions  ``hidden'' in the model-dependent hadron wave functions
or GPD parametrizations. 
An approach that is more directly connected to QCD
is based on the light-cone sum rules (LCSRs)~\cite{Braun:2001tj,Braun:2006hz}.  
This technique is attractive because in LCSRs  ``soft'' contributions to the form 
factors are calculated as an expansion in terms of the 
momentum fraction distributions of partons at small transverse separations,
dubbed distribution amplitudes (DAs), which are the same quantities 
that enter the calculation in pQCD,
and there is no double counting. 
Thus the LCSRs provide one with the most direct relation of the hadron 
form factors and DAs that is available at present, with no other nonperturbative parameters.  
Unfortunately, with the exception of the $\Delta(1232)$ resonance, up to now there exists almost no 
information on the DAs of nucleon resonances. Thus pQCD predictions 
\cite{Carlson:1985mm,Carlson:1988gt} cannot be quantified and the LCSRs cannot be
used as well. 

Moments of the DAs can, however, be calculated on the lattice.
In this work we suggest a synthetic approach combining 
the constraints on DAs from a lattice calculation with LCSRs to calculate the form factors.
As the first demonstration of this strategy we consider the electroproduction of 
$N^*(1535)$, the parity partner of the nucleon.
This is a special case because lattice calculations of baryon correlation functions
 always yield results for baryons of both parities, $J^P=1/2^+$ and $J^P=1/2^-$
(see, e.g., \cite{Lee:1998cx,Sasaki:2001nf}),
so in fact the results for $N^*(1535)$ appear to be a byproduct of our 
calculation of the nucleon DAs \cite{Gockeler:2008xv,Braun:2008ur}, 
to which we refer for further technical details.
%albeit they have somewhat larger errors.      
We find that the shapes of the nucleon and $N^*$ DAs are rather different.  
A preliminary account of this study was presented in \cite{Warkentin:2008iu}.
In this paper we further use our results on the DAs to calculate the helicity amplitudes 
$A_{1/2}(Q^2)$ and $S_{1/2}(Q^2)$ for the electroproduction of $N^*(1535)$ in the 
LCSR approach. In the region  $Q^2 > 2$~GeV$^2$, where the
light-cone expansion may be expected to converge, the results appear to
be in good agreement with the experimental data.

\phantomsection
\addcontentsline{toc}{subsubsection}{Distribution Amplitudes}
\paragraph{Distribution Amplitudes. ---}

The leading-twist(=3) nucleon (proton) DA can be defined from a 
matrix element of a nonlocal light-ray operator that involves quark 
fields of given helicity    
$q^{\up(\down)} = (1/2) (1 \pm \gamma_5) q$~\cite{Braun:2000kw}:
\begin{eqnarray}
\label{vector-twist-3}
\lefteqn{
\langle 0 | 
\epsilon^{ijk}\! \left(u^{\up}_i(a_1 n) C \!\!\not\!{n} u^{\down}_j(a_2 n)\right)  
\!\not\!{n} d^{\up}_k(a_3 n) 
|N(P)\rangle }
\nonumber\\
&=& - \frac12 f_N\,P \cdot n\! \not\!{n}\, u_N^\up(P)\! \!\int\! [dx] 
\,e^{-i P \cdot n \sum x_i a_i}\, 
\varphi_N(x_i).
\end{eqnarray}
Here $P_\mu$, $P^2=m_N^2$, is the proton momentum, $u_N(P)$ the usual Dirac spinor in 
relativistic normalization, $n_\mu$ an arbitrary light-like vector $n^2=0$ and $C$
the charge-con\-ju\-ga\-tion matrix. The variables $x_1,x_2,x_3$ have 
the meaning of the momentum fractions carried by the three valence quarks and the 
integration measure is defined as 
$\int [dx] = \int_0^1  dx_1 dx_2 dx_3 \delta(\sum x_i-1)$.
The Wilson lines that ensure gauge invariance are inserted between the quarks;
they are not shown for brevity.

The nonlocal operator 
on the l.h.s. of (\ref{vector-twist-3})
does not have a definite parity. Thus the same operator couples also 
to $N^*(1535)$ and one can define the corresponding leading-twist DA as
\begin{eqnarray}
\label{vector-twist-3-star}
\lefteqn{
\langle 0 | \epsilon^{ijk}\! 
\left(u^{\up}_i(a_1 n) C \!\!\not\!{n} u^{\down}_j(a_2 n)\right)  
\!\not\!{n} d^{\up}_k(a_3 n) |N^*(P)\rangle }
\nonumber\\
&=&  \frac12 f_{N^*}\, P \cdot n\! \not\!{n}\, u_{N^*}^{\up}(P)\! 
\!\int\! [dx] \,e^{-i P \cdot n \sum x_i a_i}\, 
\varphi_{N^*}(x_i)\,,  \nonumber
\end{eqnarray}
where, of course, $P^2=m_{N^*}^2$.
The normalization constants $f_N$ and $f_{N^*}$ are defined as
\begin{eqnarray}
&& \langle 0 |\epsilon^{ijk}\! 
\left(u_i C\slashed{n} u_j\right)\!(0)  
\!\gamma_5\slashed{n} d_k(0) |N\!(P)\rangle\! =\!f_{N} P \cdot n  \, \slashed{n}\, u_N\!(P)
\nonumber
\\
&& \langle 0 |\epsilon^{ijk}\! 
\left(u_i C\slashed{n} u_j\right)\!(0)  
\!\gamma_5\slashed{n} d_k(0) |N^*\!(P)\rangle
\nonumber \\
&& \hspace*{3.5cm}
 = f_{N^*} P \cdot n \gamma_5 \slashed{n}\, u_{N^*}\!(P) \,.
\end{eqnarray} 

On the lattice one can calculate moments of the DA
\begin{displaymath}
 \varphi^{lmn} = \int [dx]\, x_1^l x_2^m x_3^n \,\varphi(x_i) \,,
\end{displaymath}
which are related to matrix elements of local three-quark 
operators with covariant derivatives, see \cite{Braun:2008ur} for details.
The normalization is such that $\varphi^{000} = 1$.

There exist three independent subleading twist-4 
distribution amplitudes $\Phi^{N^*}_4$, $\Psi^{N^*}_4$,
$\Xi^{N^*}_4$ (as for the nucleon).
They can be defined as (cf.~\cite{Braun:2000kw,Braun:2008ia})
\begin{eqnarray}
\label{vector-twist-4.1-star}
\lefteqn{
\langle 0 | \epsilon^{ijk}\! 
\left(u^{\up}_i(a_1 n) C\slashed{n} u^{\down}_j(a_2 n)\right)  
\!\slashed{P} d^{\up}_k(a_3 n) |N^*(P)\rangle }
\nonumber\\
&=& \frac14
\,P \cdot n\, \slashed{P}\, u_{N^*}^{\up}(P)\! \!\int\! [dx] 
\,e^{-i P \cdot n \sum x_i a_i}\,
\nonumber\\
&&\times \left[f_{N^*}\Phi^{N^*,WW}_4(x_i)+\lambda^*_1\Phi^{N^*}_4(x_i)\right],   
\nonumber\\
\lefteqn{
\langle 0 | \epsilon^{ijk}\! 
\left(u^{\up}_i(a_1 n) C \slashed{n}\gamma_{\perp}\slashed{P} u^{\down}_j(a_2 n)\right)  
\gamma^{\perp}\slashed{n} d^{\up}_k(a_3 n) |N^*(P)\rangle }
\nonumber\\
&=&
-\frac12
\, P \cdot n\! \not\!{n}\,m_{N^*} u_{N^*}^{\up}(P)\! \!\int\! [dx] 
\,e^{-i P \cdot n \sum x_i a_i}\,
\nonumber\\
&&\times \left[f_{N^*}\Psi^{N^*,WW}_4(x_i)-\lambda^*_1\Psi^{N^*}_4(x_i)\right],\hspace*{2cm}\phantom{.}
\nonumber\\
\lefteqn{
\langle 0 | \epsilon^{ijk}\! 
\left(u^{\up}_i(a_1 n) C\slashed{P}\slashed{n} u^{\up}_j(a_2 n)\right)  
\!\not\!{n} d^{\up}_k(a_3 n) |N^*(P)\rangle }
\nonumber\\
&=& \frac{\lambda^*_2}{12}\, P \cdot n\! \not\!{n}\, m_{N^*} u_{N^*}^{\up}(P)\! \!\int\! [dx] 
\,e^{-i P \cdot n \sum x_i a_i}\,\Xi^{N^*}_4(x_i) \,,    \nonumber
\end{eqnarray}
where $\Phi^{N^*,WW}_4(x_i)$ and $\Psi^{N^*,WW}_4(x_i)$ are the so-called 
Wandzura-Wilczek contributions, which can be expressed in terms of the 
leading-twist DA \cite{Braun:2008ia}. The two new normalization 
constants are given by the local matrix elements
\begin{eqnarray}
&& \langle 0 | \epsilon^{ijk}\! 
\left(u_i C\gamma_{\mu} u_j\right)\!(0)  
\!\gamma_5\gamma^{\mu} d_k(0) |N^*\!(P)\rangle
\nonumber \\
&& \hspace*{4.0cm} = \lambda_1^* m_{N^*} \gamma_5 u_{N^*}\!(P),
\nonumber
\\
&& \langle 0 | \epsilon^{ijk}\! 
\left(u_i C\sigma_{\mu\nu} u_j\right)\!(0)  
\!\gamma_5\sigma^{\mu \nu} d_k(0) |N^*\!(P)\rangle
\nonumber \\ 
&& \hspace*{4.0cm} = \lambda_2^{*} m_{N^*} \gamma_5 u_{N^*}\!(P).
\nonumber
\end{eqnarray}

\begin{table}[t]
 \begin{center}  
\renewcommand{\arraystretch}{1.15}
\vspace{2ex}
\centering
\begin{tabular}{|c|c|c|c|c|c|c|}
\hline
  & Asympt.& $N$ & $N^\star (1535)$ \\ 
\hline
$ f_N$      &      &  3.234(63)(86)  & 4.544(117)(223)  \\
$-\lambda_1$&      & 35.57(65)(136)  & 37.55(101)(768)  \\
$ \lambda_2$&      & 70.02(128)(268) & 191.9(44)(391)   \\
\hline
  $\varphi^{100}$ & $\frac13\simeq 0.333$      & 0.3999(37)(139)  & 0.4765(33)(155)   \\
  $\varphi^{010}$ & $\frac13\simeq 0.333$      & 0.2986(11)(52)   & 0.2523(20)(32)    \\
  $\varphi^{001}$ & $\frac13\simeq 0.333$      & 0.3015(32)(106)  & 0.2712(41)(136)   \\
\hline
  $\varphi^{200}$ & $\frac17\simeq 0.143$      & 0.1816(64)(212)  & 0.2274(89)(307)   \\
  $\varphi^{020}$ & $\frac17\simeq 0.143$      & 0.1281(32)(106)  & 0.0915(45)(224)   \\
  $\varphi^{002}$ & $\frac17\simeq 0.143$      & 0.1311(113)(382) & 0.1034(160)(584)  \\
  $\varphi^{011}$ & $\frac{2}{21}\simeq 0.095$ & 0.0613(89)(319)  & 0.0398(132)(497)  \\
  $\varphi^{101}$ & $\frac{2}{21}\simeq 0.095$ & 0.1091(41)(152)  & 0.1281(56)(131)   \\
  $\varphi^{110}$ & $\frac{2}{21}\simeq 0.095$ & 0.1092(67)(219)  & 0.1210(101)(304)  \\
\hline                                                                           
\end{tabular}
\end{center}                                                                     
\caption{The normalization constants (in units of $10^{-3}\,\mathrm{GeV}^2$)
and moments of the DAs obtained from 
{QCDSF/DIK} configurations at $\beta=5.40$ for the nucleon ($N$) 
and $N^\star(1535)$ at $\mu^2_{\overline{MS}}=1\,\mathrm{GeV}^2$. 
The first error is statistical, the second error represents the uncertainty 
due to the chiral extrapolation and renormalization. The systematic error should be 
considered with caution.
}
\label{tab:1}
\end{table}

The asymptotic distribution amplitudes (at very large scales) for the 
nucleon and $N^*$ are the same: 
\begin{eqnarray}
 &&\phi^{\rm as}(x_i) = 120 x_1 x_2 x_3\,,\quad \Phi^{\rm as}_4(x_i) =  24
x_1 x_2\,,\quad
\nonumber\\
&& \Phi^{WW,{\rm as}}_4(x_i)= 24 x_1 x_2 (1+\frac{2}{3}(1-5x_3))\,,
\nonumber\\
&&\Psi^{WW,{\rm as}}_4(x_i)= 24 x_1 x_3(1+\frac{2}{3}(1-5x_2))\,,
\nonumber\\
&& \Xi_4(x_i) = 24 x_2 x_3\,,\quad  \Psi^{\rm as}_4(x_i) = 24 x_1 x_3\,.
\nonumber
\end{eqnarray}

Baryon states of different parity can be identified in a lattice calculation as those propagating
forward or backward in (imaginary) time as long as their momentum vanishes
\cite{Lee:1998cx,Sasaki:2001nf}. While vanishing momentum is sufficient for the evaluation of the
normalization constants, the higher moments of the DAs require nonzero momentum. In this case the
signal in the negative parity channel is contaminated by a contribution of the $J^P=1/2^+$ (nucleon)
ground state of the order $\vec{p}^2/(m_N m_{N^*})$ enhanced by the factor $e^{(E_{N^*} -
E_N)(T-t)}$, where $T$ is the time extent of our lattice. However, this effect seems to be quite
small in our results: Replacing the parity projector $ (1/2) \left( 1 + \gamma_4 \right)$ by $ (1/2)
\left( 1 + (m_N/E_N) \gamma_4 \right)$ \cite{Lee:1998cx} changes the first (second) moments of the
$N^*$ DAs by 1\% (5\%), which is well below the statistical error. In principle, there is still a
contamination by the $J^P=1/2^+$ $N^*(1440)$ (Roper) resonance, but for small momenta this effect is
expected to be negligible \cite{Lee:1998cx}. Another issue is that in the physical spectrum there
are two $J^P=1/2^-$ resonances, $N^*(1535)$ and $N^*(1650)$, which are close to each other, so that
they cannot be distinguished by means of their mass difference in our calculation. Because of the
peculiar decay pattern of $N^*(1650)$ we expect, however, that this state has a much smaller
coupling to the usual interpolating operators \cite{Sasaki:2001nf}. So our results can be identified
with the contribution of $N^*(1535)$ alone. All these questions certainly deserve a further study.

The results of our calculation of the normalization constants $f_{N^*}$, $\lambda_1^*$,  $\lambda_2^*$ 
and of the first few moments of the leading-twist DA of the $N^*(1535)$ resonance are compared to the similar
calculation for the nucleon \cite{Braun:2008ur} in Table~\ref{tab:1}. It attracts attention that 
$f_{N^*}$ is about 50\% larger than $f_N$. 
This means that the wave function of the three quarks at the origin is
larger in the $J^P=1^-$ state than in the $J^P=1^+$ state, which may be 
counterintuitive.
The momentum fraction carried by the u-quark with the same helicity as the baryon itself, $\varphi^{100}$,
appears to be considerably larger for $N^*$, indicating that its DA is more asymmetric.
These features suggest that the large asymmetry of the nucleon DA observed in 
QCD sum rule calculations \cite{Chernyak:1984bm,King:1986wi,Chernyak:1987nu} 
may be due to a contamination of the sum rules by states of opposite parity, 
which are difficult to separate in this approach.

\begin{figure}[t]
\includegraphics[width=0.47\textwidth,clip]{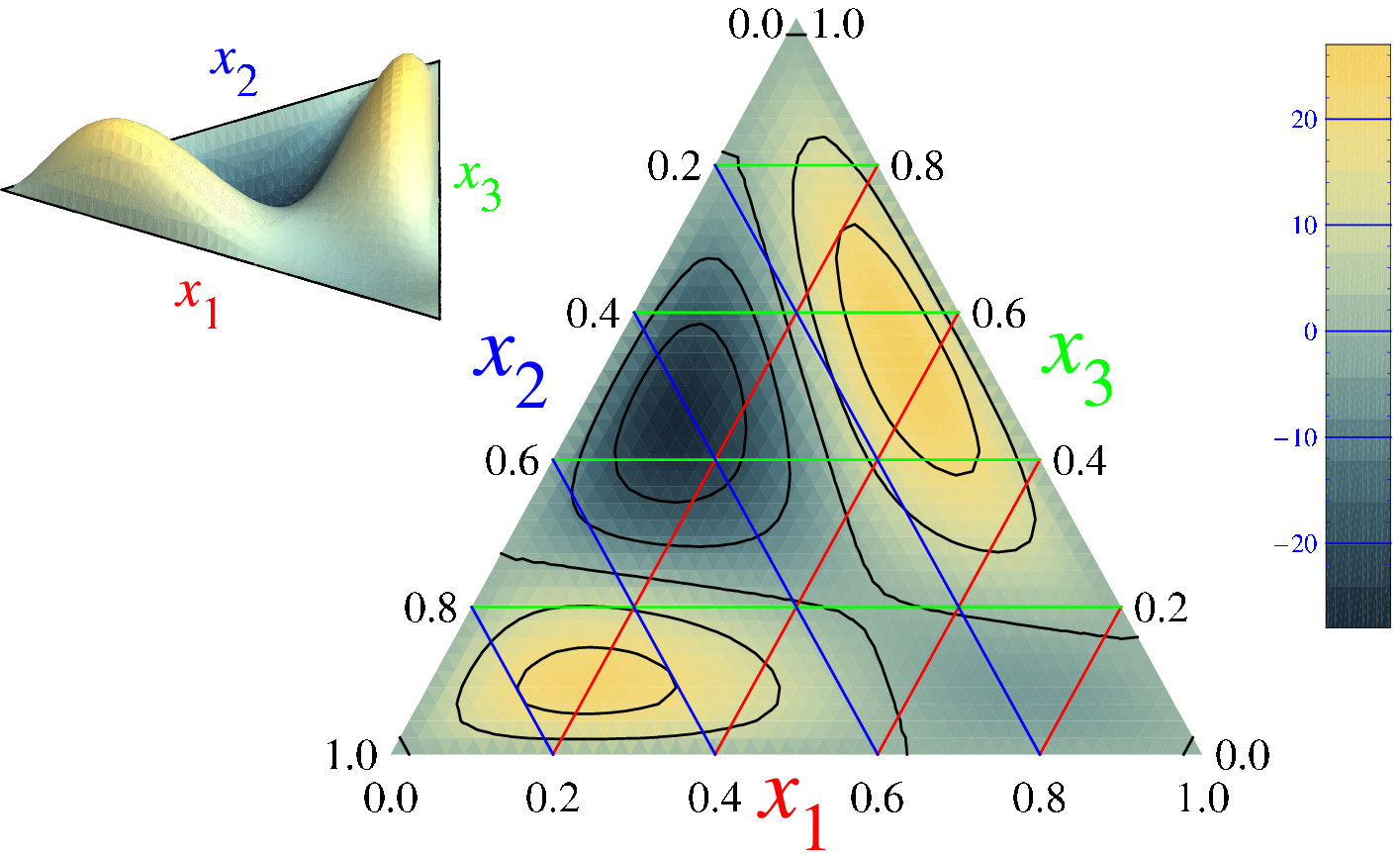}\\[1mm]
\includegraphics[width=0.47\textwidth,clip]{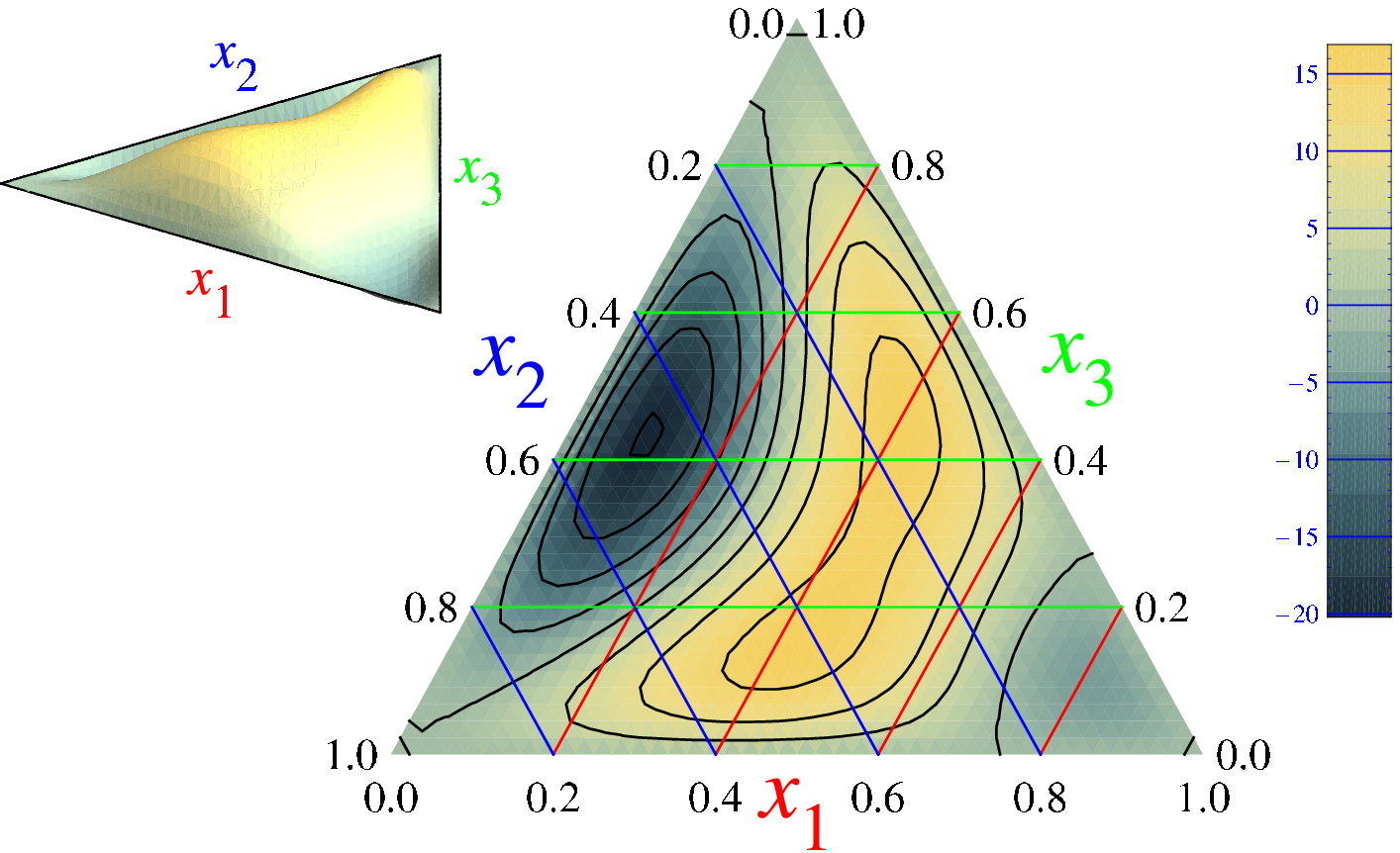}
\caption{
Barycentric plot of the distribution amplitudes for nucleon (up) and $N^\star(1535)$ (down) at 
$\mu_{\overline{MS}}=1\mathrm{GeV}$ using the central values of the lattice results.
The lines of constant $x_1$, $x_2$ and $x_3$ are parallel 
to the sides of the triangle labelled by $x_2$, $x_3$ and $x_1$, respectively.
}
\label{ndapic}
\end{figure}

The calculated moments can be used to model the $N^*$ leading-twist DA as an 
expansion in orthogonal polynomials corresponding to the contributions 
of multiplicatively renormalizable operators (in leading order), see \cite{Braun:2008ur}.  
The comparison of such models for $N$ and $N^*$, obtained using the polynomial expansion 
to second order and the central values of the lattice parameters, is shown in 
Fig.~\ref{ndapic}.

\phantomsection
\addcontentsline{toc}{subsubsection}{Helicity Amplitudes from LCSRs}
\paragraph{Helicity Amplitudes from LCSRs. ---}
The matrix element of the electromagnetic current $j^{\rm em}_\nu$ between spin-1/2
states of opposite parity can be parametrized in terms of two independent 
form factors, which can be chosen as
\begin{eqnarray} 
\label{Nstar_FF}
&&\langle N^*(P') | j_{\nu}^{\rm em} |N(P)\rangle \,=\, 
 \bar{u}_{N^*}(P')\gamma_5 \Gamma_\nu u_N(P)\,,
\nonumber\\
&& \Gamma_\nu \,=\,\frac{{G_1(q^2)}}{m_N^2}(\slashed{q} q_\nu - q^2 \gamma_\nu)
 -i \frac{{G_2(q^2)}}{m_N} \sigma_{\nu\rho} q^{\rho} \,,
\end{eqnarray}
where $q=P'-P$ is the momentum transfer. The helicity amplitudes 
$A_{1/2}(Q^2)$ and $S_{1/2}(Q^2)$ for the electroproduction of $N^*(1535)$
can be expressed in terms of these form factors \cite{Aznauryan:2008us}:
\begin{eqnarray}
A_{1/2} &=& e\, B\, 
\Big[ Q^2 G_1(Q^2) + m_N(m_{N^*}-m_N)G_2(Q^2) \Big],
\nonumber \\
{S}_{1/2}&=&  \frac{e}{\sqrt{2}}B\,C\, 
\Big[(m_{N}-m_{N^*})G_1(Q^2)+m_{N}G_2(Q^2)\Big].
\nonumber
\end{eqnarray}
Here $e$ is the elementary charge and $B$, $C$ are kinematic factors defined as
\begin{eqnarray}
   B &=& \sqrt{\frac{Q^2+(m_{N*}+m_N)^2}{2 m_N^5(m_{N^*}^2-m_N^2)}} \,,
\nonumber\\ 
   C & =& \sqrt{1+\frac{(Q^2-m_{N^*}^2+m_N^2)^2}{4 Q^2 m_{N^*}^2}} \,.
\nonumber
\end{eqnarray}

The LCSRs are derived from the correlation function 
\begin{displaymath}
\int\! dx\, e^{-iqx}\langle N^*(P)| T \{ \eta (0) j_\mu^{\rm em} (x) \} 
| 0 \rangle \,,
\end{displaymath}
where $\eta$ is a suitable operator with nucleon quantum numbers, e.g. 
the Ioffe current~\cite{Ioffe:1981kw}.
Making  use of the duality of QCD quark-gluon and hadronic degrees of freedom  
through dispersion relations one can write a representation for the 
form factors appearing in (\ref{Nstar_FF}) in terms of the DAs of $N^*$. 
In leading order, the  sum rules for $Q^2G_1(Q^2)/(m_N m_{N^*})$ and $-2G_2(Q^2)$
have the same functional form as the similar sum rules~\cite{Braun:2001tj,Braun:2006hz} 
for the Dirac and Pauli electromagnetic form factors of the proton,
with the replacement $m_N\to m_{N^*}$ in the light cone expansion part, 
and different DAs. 
%only the nonperturbatuve input being different.  

\begin{figure}[t]
\includegraphics[width=0.45\textwidth,clip]{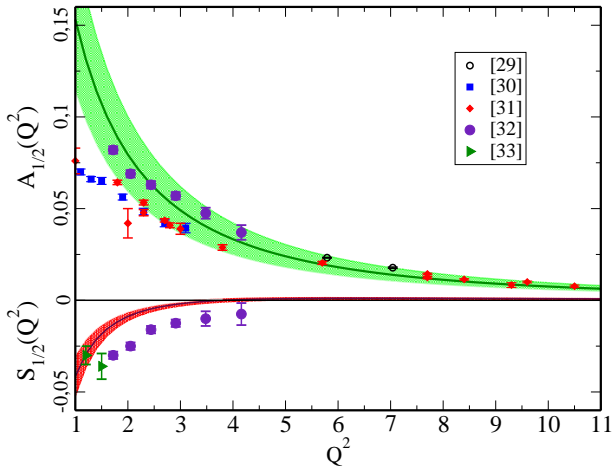}
\caption{The LCSR calculation for the helicity amplitudes
$A_{1/2}(Q^2)$ and $S_{1/2}(Q^2)$ for the electroproduction of the 
$N^*(1535)$ resonance using the lattice results from Table~\ref{tab:1} 
for the lowest moments of the  $N^*(1535)$ DAs. 
The curves are obtained using the central values of the lattice 
parameters, and the shaded areas show the corresponding uncertainty.}
\label{fig:A12S12}
\end{figure}

In the present calculation we used a model for the leading-twist DA 
including first order corrections in the polynomial expansion, asymptotic 
expressions for the ``genuine'' twist-4 DAs and the corresponding Wandzura-Wilczek
corrections up to twist-6 as given in Ref.~\cite{Braun:2000kw}.  
The results are shown in Fig.~\ref{fig:A12S12}. The shaded areas correspond to the 
uncertainty in the lattice values as given in Table~\ref{tab:1}. 
In the region  $Q^2 > 2$~GeV$^2$, where the
light-cone expansion may be expected to converge, the results appear to
be in good agreement with the experimental data.

\phantomsection
\addcontentsline{toc}{subsubsection}{Discussion and Conclusions.}
\paragraph{Discussion and Conclusions. ---}

In this work we suggest to calculate transition form factors for nucleon
resonances at intermediate momentum tranfer, combining
the constraints on DAs from a lattice calculation with LCSRs 
This approach seems to be especially promising for $N^*(1535)$,
the parity partner of the nucleon, because of the relative ease to separate the states 
of opposite parity on the lattice. The accuracy is expected to increase 
significantly when calculations with smaller pion masses and on larger lattices 
become available. This would remove a major source of uncertainties
which is due to the chiral extrapolation. 

In order to match the accuracy of the lattice results, the LCSR calculations of baryon form 
factors will have to be advanced to include NLO radiative corrections, as it has become 
standard for meson decays. For the first effort in this direction see
\cite{PassekKumericki:2008sj}. In addition, one needs a technique for the 
resummation of ``kinematic'' corrections to the sum rules 
that are due to the large masses of the resonances.

\paragraph*{Acknowledgment. ---}
\begin{acknowledgments}
We are grateful to A.~Afanasev, I.~Aznauryan  and A.~Manashov for helpful discussions. 
The numerical calculations have been performed on the Hitachi
SR8000 at LRZ (Munich), apeNEXT and APEmille at NIC/DESY (Zeuthen) and
BlueGene/Ls at NIC/JSC (J\"ulich), EPCC (Edinburgh) and KEK (by the Kanazawa
group as part of the DIK research program) as well as  QCDOC (Regensburg)
 using the Chroma software library \cite{Edwards:2004sx, bagel:2005}. 
This work was supported by DFG (Forschergruppe 
``Gitter-Hadronen-Ph\"anomenologie'', grant 9209070 and 
SFB/TR 55 ``Hadron Physics from Lattice QCD''), 
by EU I3HP (contract No. RII3-CT-2004-506078)  and by BMBF.
\end{acknowledgments}

\providecommand{\href}[2]{#2}\begingroup\raggedright
\endgroup


\begin{thebibliography}{99}

%\cite{Chernyak:1977as}
\bibitem{Chernyak:1977as}
  V.~L.~Chernyak and A.~R.~Zhitnitsky,
  %``Asymptotic Behavior Of Hadron Form-Factors In Quark Model. (In Russian),''
  JETP Lett.\  {\bf 25}, 510 (1977).
  %[Pisma Zh.\ Eksp.\ Teor.\ Fiz.\  {\bf 25}, 544 (1977)].
  %%CITATION = ZFPRA,25,544;%%

%\cite{Chernyak:1977fk}
\bibitem{Chernyak:1977fk}
  V.~L.~Chernyak, A.~R.~Zhitnitsky and V.~G.~Serbo,
  %``Asymptotic hadronic form-factors in quantum chromodynamics,''
  JETP Lett.\  {\bf 26}, 594 (1977).
  %[Pisma Zh.\ Eksp.\ Teor.\ Fiz.\  {\bf 26}, 760 (1977)].
  %%CITATION = ZFPRA,26,760;%%

%\cite{Efremov:1979qk}
\bibitem{Efremov:1979qk}
  A.~V.~Efremov and A.~V.~Radyushkin,
  %``Factorization And Asymptotical Behavior Of Pion Form-Factor In QCD,''
  Phys.\ Lett.\  B {\bf 94}, 245 (1980).
  %%CITATION = PHLTA,B94,245;%%

%\cite{Efremov:1978rn}
\bibitem{Efremov:1978rn}
  A.~V.~Efremov and A.~V.~Radyushkin,
  %``Asymptotical Behavior Of Pion Electromagnetic Form-Factor In QCD,''
  Theor.\ Math.\ Phys.\  {\bf 42}, 97 (1980).
  %[Teor.\ Mat.\ Fiz.\  {\bf 42}, 147 (1980)].
  %%CITATION = TMFZA,42,147;%%

%\cite{Lepage:1979za}
\bibitem{Lepage:1979za}
  G.~P.~Lepage and S.~J.~Brodsky,
  %``Exclusive Processes In Quantum Chromodynamics: The Form-Factors Of Baryons
  %At Large Momentum Transfer,''
  Phys.\ Rev.\ Lett.\  {\bf 43}, 545 (1979)
  [Erratum-ibid.\  {\bf 43}, 1625 (1979)].
  %%CITATION = PRLTA,43,545;%%

%\cite{Lepage:1980fj}
\bibitem{Lepage:1980fj}
  G.~P.~Lepage and S.~J.~Brodsky,
  %``Exclusive Processes In Perturbative Quantum Chromodynamics,''
  Phys.\ Rev.\  D {\bf 22}, 2157 (1980).
  %%CITATION = PHRVA,D22,2157;%%

%\cite{Burkert:2008mw}
\bibitem{Burkert:2008mw}
  V.~D.~Burkert,
  %``Electromagnetic Transition Form Factors of Nucleon Resonances,''
  AIP Conf.\ Proc.\  {\bf 1056}, 348 (2008)
  [arXiv:0808.2326 [nucl-ex]].
  %%CITATION = APCPC,1056,348;%%

\bibitem{Nstar_workshop}
%%see e.g. presentations at 
%``Electromagnetic N-N* Transition Form Factors Workshop,''
%October 13-15, 2008, Jefferson Lab, Newport News, USA;~
%{\it http://conferences.jlab.org/EmNN/program.html}
  R. Gothe, V. Mokeev {\it et al.},
  Research Proposal ``Nucleon Resonance Studies with CLAS12'',
  \# PR-09-003, approved by PAC34

   %\cite{Kroll:1995pv}
\bibitem{Kroll:1995pv}
  P.~Kroll, M.~Schurmann and P.~A.~M.~Guichon,
  %``Virtual Compton scattering off protons at moderately large momentum
  %transfer,''
  Nucl.\ Phys.\ A {\bf 598}, 435 (1996).
  %[arXiv:hep-ph/9507298].
  %%CITATION = HEP-PH 9507298;%%

%\cite{Bolz:1996sw}
\bibitem{Bolz:1996sw}
  J.~Bolz and P.~Kroll,
  %``Modelling the nucleon wave function from soft and hard processes,''
  Z.\ Phys.\  A {\bf 356}, 327 (1996).
  %[arXiv:hep-ph/9603289].
  %%CITATION = ZEPYA,A356,327;%%

  %\cite{Belitsky:2003nz}
\bibitem{Belitsky:2003nz}
  A.~V.~Belitsky, X.~Ji and F.~Yuan,
  %``Quark imaging in the proton via quantum phase-space distributions,''
  Phys.\ Rev.\ D {\bf 69}, 074014 (2004).
  %[arXiv:hep-ph/0307383].
  %%CITATION = HEP-PH 0307383;%%
  
%\cite{Diehl:2004cx}
\bibitem{Diehl:2004cx}
  M.~Diehl, T.~Feldmann, R.~Jakob and P.~Kroll,
  %``Generalized parton distributions from nucleon form factor data,''
  Eur.\ Phys.\ J.\ C {\bf 39}, 1 (2005).
  %[arXiv:hep-ph/0408173].
  %%CITATION = HEP-PH 0408173;%%
 
 %\cite{Guidal:2004nd}
\bibitem{Guidal:2004nd}
M.~Guidal, M.~V.~Polyakov, A.~V.~Radyushkin and M.~Vanderhaeghen,
%``Nucleon form factors from generalized parton distributions,''
Phys.\ Rev.\ D {\bf 72}, 054013 (2005).
%[arXiv:hep-ph/0410251].
%%CITATION = HEP-PH 0410251;%%

%\cite{Braun:2001tj}
\bibitem{Braun:2001tj}
  V.~M.~Braun, A.~Lenz, N.~Mahnke and E.~Stein,
  %``Light-cone sum rules for the nucleon form factors,''
  Phys.\ Rev.\  D {\bf 65}, 074011 (2002).
  %[arXiv:hep-ph/0112085].
  %%CITATION = PHRVA,D65,074011;%%

%\cite{Braun:2006hz}
\bibitem{Braun:2006hz}
  V.~M.~Braun, A.~Lenz and M.~Wittmann,
  %``Nucleon form factors in QCD,''
  Phys.\ Rev.\  D {\bf 73}, 094019 (2006).
  %[arXiv:hep-ph/0604050].
  %%CITATION = PHRVA,D73,094019;%%

%\cite{Carlson:1985mm}
\bibitem{Carlson:1985mm}
  C.~E.~Carlson,
  %``Electromagnetic N - Delta Transition At High Q**2,''
  Phys.\ Rev.\  D {\bf 34}, 2704 (1986).
  %%CITATION = PHRVA,D34,2704;%%

%\cite{Carlson:1988gt}
\bibitem{Carlson:1988gt}
  C.~E.~Carlson and J.~L.~Poor,
  %``DISTRIBUTION AMPLITUDES AND ELECTROPRODUCTION OF THE DELTA AND OTHER LOW
  %LYING RESONANCES,''
  Phys.\ Rev.\  D {\bf 38}, 2758 (1988).
  %%CITATION = PHRVA,D38,2758;%%

%\cite{Lee:1998cx}
\bibitem{Lee:1998cx}
  F.~X.~Lee and D.~B.~Leinweber,
  %``Negative-parity baryon spectroscopy,''
  Nucl.\ Phys.\ Proc.\ Suppl.\  {\bf 73}, 258 (1999).
  %[arXiv:hep-lat/9809095].
  %%CITATION = NUPHZ,73,258;%%

%\cite{Sasaki:2001nf}
\bibitem{Sasaki:2001nf}
  S.~Sasaki, T.~Blum and S.~Ohta,
  %``A lattice study of the nucleon excited states with domain wall  fermions,''
  Phys.\ Rev.\  D {\bf 65}, 074503 (2002).
  %[arXiv:hep-lat/0102010].
  %%CITATION = PHRVA,D65,074503;%%


%\cite{Gockeler:2008xv}
\bibitem{Gockeler:2008xv}
  M.~G\"ockeler {\it et al.},
  %``Nucleon distribution amplitudes from lattice QCD,''
  Phys.\ Rev.\ Lett.\  {\bf 101}, 112002 (2008).
  %[arXiv:0804.1877 [hep-lat]].
  %%CITATION = PRLTA,101,112002;%%

%\cite{Braun:2008ur}
\bibitem{Braun:2008ur}
  V.~M.~Braun {\it et al.},
  %``Nucleon distribution amplitudes and proton decay matrix elements on the
  %lattice,''
  arXiv:0811.2712 [hep-lat].
  %%CITATION = ARXIV:0811.2712;%%

%\cite{Warkentin:2008iu}
\bibitem{Warkentin:2008iu}
  N.~Warkentin {\it et al.},
  %``Wave functions of the nucleon and its parity partner from lattice QCD,''
  arXiv:0811.2212 [hep-lat].
  %%CITATION = ARXIV:0811.2212;%%

%\cite{Braun:2000kw}
\bibitem{Braun:2000kw}
  V.~Braun, R.~J.~Fries, N.~Mahnke and E.~Stein,
  %``Higher twist distribution amplitudes of the nucleon in QCD,''
  Nucl.\ Phys.\  B {\bf 589}, 381 (2000)
  [Erratum-ibid.\  B {\bf 607}, 433 (2001)].
  %[arXiv:hep-ph/0007279].
  %%CITATION = NUPHA,B589,381;%%

%\cite{Braun:2008ia}
\bibitem{Braun:2008ia}
  V.~M.~Braun, A.~N.~Manashov and J.~Rohrwild,
  %``Baryon Operators of Higher Twist in QCD and Nucleon Distribution
  %Amplitudes,''
  Nucl.\ Phys.\  B {\bf 807}, 89 (2009).
  %[arXiv:0806.2531 [hep-ph]].
  %%CITATION = NUPHA,B807,89;%%

%\cite{Chernyak:1984bm}
\bibitem{Chernyak:1984bm}
  V.~L.~Chernyak and I.~R.~Zhitnitsky,
  %``Nucleon Wave Function And Nucleon Form-Factors In QCD,''
  Nucl.\ Phys.\  B {\bf 246}, 52 (1984).
  %%CITATION = NUPHA,B246,52;%%

%\cite{King:1986wi}
\bibitem{King:1986wi}
  I.~D.~King and C.~T.~Sachrajda,
  %``Nucleon Wave Functions and QCD Sum Rules,''
  Nucl.\ Phys.\  B {\bf 279}, 785 (1987).
  %%CITATION = NUPHA,B279,785;%%

%\cite{Chernyak:1987nu}
\bibitem{Chernyak:1987nu}
  V.~L.~Chernyak, A.~A.~Ogloblin and I.~R.~Zhitnitsky,
  %``The wave functions of the octet baryons,''
  Z.\ Phys.\  C {\bf 42}, 569 (1989).
  %[Yad.\ Fiz.\  {\bf 48}, 1410 (1988\ SJNCA,48,896-904.1988)].
  %%CITATION = SJNCA,48,896;%%

%\cite{Aznauryan:2008us}
\bibitem{Aznauryan:2008us}
  I.~G.~Aznauryan, V.~D.~Burkert and T.~S.~Lee,
  %``On the definitions of the gamma*N -> N* helicity amplitudes,''
  arXiv:0810.0997 [nucl-th].
  %%CITATION = ARXIV:0810.0997;%%

\bibitem{Dalton:2008ff}
  M.~M.~Dalton {\it et al.},
  %``Electroproduction of Eta Mesons in the S11(1535) Resonance Region at High
  %Momentum Transfer,''
  arXiv:0804.3509 [hep-ex].
  %%CITATION = ARXIV:0804.3509;%%


\bibitem{Denizli:2007tq}
  H.~Denizli {\it et al.}  [CLAS Collaboration],
  %``Q^2 Dependence of the S_{11}(1535) Photocoupling and Evidence for a
  %P-wave resonance in eta electroproduction,''
  Phys.\ Rev.\  C {\bf 76}, 015204 (2007).
  %[arXiv:0704.2546 [nucl-ex]].
  %%CITATION = PHRVA,C76,015204;%%

\bibitem{Stoler:1993yk}
  P.~Stoler,
  %``Baryon form-factors at high Q**2 and the transition to perturbative QCD,''
  Phys.\ Rept.\  {\bf 226}, 103 (1993);\\
  %%CITATION = PRPLC,226,103;%%
  F.~W.~Brasse, W.~Flauger, J.~Gayler, S.~P.~Goel, R.~Haidan, M.~Merkwitz and H.~Wriedt,
  %``Parametrization Of The Q**2 Dependence Of Virtual Gamma P Total
  %Cross-Sections In The Resonance Region,''
  Nucl.\ Phys.\  B {\bf 110}, 413 (1976).
  %%CITATION = NUPHA,B110,413;%%

\bibitem{AznPriCom}
I.~G.~Aznauryan, private communication, see also \cite{Nstar_workshop}.

\bibitem{Tiator:2003uu}
  L.~Tiator, D.~Drechsel, S.~Kamalov, M.~M.~Giannini, E.~Santopinto and A.~Vassallo,
  %``Electroproduction of nucleon resonances,''
  Eur.\ Phys.\ J.\  A {\bf 19}, 55 (2004)
  %[arXiv:nucl-th/0310041].
  %%CITATION = EPHJA,A19,55;%%

%\cite{Ioffe:1981kw}
\bibitem{Ioffe:1981kw}
  B.~L.~Ioffe,
  %``Calculation Of Baryon Masses In Quantum Chromodynamics,''
  Nucl.\ Phys.\  B {\bf 188}, 317 (1981)
  [Erratum-ibid.\  B {\bf 191}, 591 (1981)].
  %%CITATION = NUPHA,B188,317;%%

%\cite{PassekKumericki:2008sj}
\bibitem{PassekKumericki:2008sj}
  K.~Passek-Kumericki and G.~Peters,
  %``Nucleon Form Factors to Next-to-Leading Order with Light-Cone Sum Rules,''
  Phys.\ Rev.\  D {\bf 78}, 033009 (2008).
  %[arXiv:0805.1758 [hep-ph]].
  %%CITATION = PHRVA,D78,033009;%%

%\cite{Edwards:2004sx}
\bibitem{Edwards:2004sx}
  R.~G.~Edwards and B.~Jo\'o  [SciDAC Collaboration],
  %``The Chroma software system for lattice QCD,''
  Nucl.\ Phys.\ Proc.\ Suppl.\  {\bf 140}, 832 (2005).
  %[arXiv:hep-lat/0409003].
  %%CITATION = NUPHZ,140,832;%%

\bibitem{bagel:2005}
P.~A. Boyle (2005)
{\it http://www.ph.ed.ac.uk/~paboyle/bagel/Bagel.html}.
%\newblock \url{http://www.ph.ed.ac.uk/~paboyle/bagel/Bagel.html}.

\end{thebibliography}
\end{document}